# Space-Time Nonlocal Metasurfaces for Event-Based Image Processing

Sedigheh Esfahani[1,2,†], Michele Cotrufo[1,3,†], and Andrea Alù[1,2,*]

[1]*Photonics Initiative, Advanced Science Research Center, City University of New York, New York, NY 10031, USA*
[2]*Physics Program, Graduate Center of the City University of New York, New York, NY 10016, USA*
[3]*The Institute of Optics, University of Rochester, Rochester, New York 14627, USA*
[†] *These authors contributed equally*
*\*aalu@gc.cuny.edu*

*Analog computation with passive optical components can enhance processing speeds and reduce power consumption, recently attracting renewed interest thanks to the opportunities enabled by metasurfaces. Basic image processing tasks, such as spatial differentiation, have been recently demonstrated based on engineered nonlocalities in metasurfaces, but next-generation computational schemes require more advanced capabilities. Here, we tailor nonlocalities in space and time to design a metasurface that can perform mixed spatio-temporal differentiation of an input image, realizing event-based edge detection with a passive ultrathin silicon-based structured film compatible with standard fabrication techniques. The metasurface detects the object edges only when the object moves, and its design can be tailored to selectively enhance objects moving at desired speeds. Our results point towards fully-passive processing of spatio-temporal signals, for highly compact neuromorphic cameras.*

Designing faster and energy-efficient tools for data processing is of paramount importance due to the exponentially growing rate of data creation, processing and storage[1]. Replacing digital processing with analog optical computing has recently attracted renewed interest[2–5] due to the appealing possibility of manipulating data at the speed of light while avoiding analog-to-digital conversion[6]. Image processing is among the most important computational tasks, with critical applications in computer vision[7], neuromorphic computing[8,9], artificial intelligence[10] and augmented reality[11].

In the last decade, analog passive computing has evolved from bulky elements to ultra-thin structured layers – known as metasurfaces[12] – which are compatible with large-area fabrication and prone to miniaturization and integration[13]. Several works[14–22] have demonstrated that optical metasurfaces can perform different image processing tasks, such as edge detection[14–19,23], *without requiring a 4F system*[24,25], with the potential of drastically reducing the footprint of all-optical analog computers. In this approach, a metasurface with tailored angle-dependent transmission acts as a spatial-momentum filter, thus performing a desired spatial operation – e.g., differentiation – by filtering the spatial Fourier components of the input image. In this context, *nonlocal* metasurfaces[14–16,26–28] (whose optical response is dictated by the coherent interaction between many unit cells) supporting engineered Fano resonances and/or quasi-bound states in the continuum have been instrumental to tailor nontrivial angle-dependent responses, leading to efficient image-processing devices.

While most studies on metamaterial-based analog computation have so far focused on *spatial* operations, comparatively little effort has been devoted to the implementation of *temporal* and, more broadly, *spatio-temporal* computation. Similarly to 4F systems for spatial operations, temporal operations can be performed via bulky pulse shapers, whereby the frequency components of a pulse are separated by diffraction gratings and filtered by masks[29,30]. Notably, the working principles of metamaterial-based spatial image processing can be extended to the temporal domain: by tailoring its *frequency* response, a metasurface can act as a compact spectral filter, thus performing time-domain computations *without* using pulse shapers and diffraction gratings. Following this strategy, previous theoretical works have investigated temporal processing with graphene[31] or dielectric[32] structures, including a recent experimental demonstration at radio frequencies[33]. Recent studies have also investigated metasurfaces that perform linear combinations of spatial and temporal differentiations[34–37], to generate spatio-temporal optical vortices[38,39].

Besides signal processing, performing analog and high-speed spatio-temporal computations can largely benefit[40] the emerging field of neuromorphic computing (NC)[8,9] – a computational framework where algorithms and hardware mimic the human brain. Rather than capturing information at regular intervals, *neuromorphic* or *event-based* cameras respond to local changes of the input signal in time[41]. Current implementations of neuromorphic cameras rely on sophisticated circuitry that detects changes in brightness between neighboring pixels[42,43], triggering data acquisition. These approaches are affected by limited temporal and spatial resolution and large latency, and they require active components and voltage bias, preventing miniaturization and energy savings. Notably, the basic neuromorphic principle – a spatial computation triggered by a temporal change – can be mimicked by *mixed* spatio-temporal differentiation [e.g., $f(x,y,t) \to \partial_t^2 \nabla_{x,y}^2 f(x,y,t)$], as outlined in this Letter. Yet, implementing these operations with analog metasurfaces is challenging, as it requires engineering a highly nontrivial frequency-momentum transfer function.

In this work, we show that dispersion engineering of nonlocal metasurfaces can be extended to the spatio-temporal domain, simultaneously tailoring angular and frequency responses. In particular, we demonstrate a general recipe to design metasurfaces performing cascaded spatio-temporal differentiation. Based on this principle, we design ultrathin passive metasurfaces, with a relatively simple geometry, performing spatio-temporal *mixed* second-order differentiation [e.g., $\partial_t^2 \nabla_{x,y}^2$] on an input signal. This operation performs analog, all-optical and fully-passive *event-based edge detection*: only regions of the input signal with simultaneous nonzero spatial and time gradients are transmitted. In other words, the spatial edges of an input image are selectively transmitted *only when their local intensity changes in time*, as sketched in Fig. 1: a spatio-temporal signal f(x, y, t), encoded in the envelope of a travelling electromagnetic wave, impinges on a metasurface with tailored spatio-temporal nonlocality. For graphical simplicity, in Fig. 1 we consider a simplified case of a factorized input signal, i.e., $f(x, y, t) = f_t(t) \times f_{x,y}(x, y)$, where $f_{x,y}(x, y)$ is a binary-valued 2D image (the red rhombus), and the overall intensity changes in time following a pulse-like function $f_t(t)$ (top-left of Fig. 1). The metasurface filters this signal, creating an output signal proportional to $\partial_t^2 \nabla_{x,y}^2 f(x, y, t)$, whereby the spatial edges of the input are enhanced only at times when the input intensity is simultaneously varying. We emphasize that such functionality can only be implemented with metasurfaces performing mixed spatio-temporal operations [e.g., $\partial_t^2 \nabla_{x,y}^2$], and it cannot be achieved with devices performing linear combinations of temporal and spatial derivatives[34–37]. We numerically demonstrate that this advanced computational task can be achieved in a simple, fully passive, dielectric subwavelength metasurface. Compared to event-based electronic NCs, our approach is not affected by finite temporal resolution and latency, and it involves no energy consumption. Moreover, the sub-wavelength footprint of the proposed metasurface allows for easy integration in compact devices.

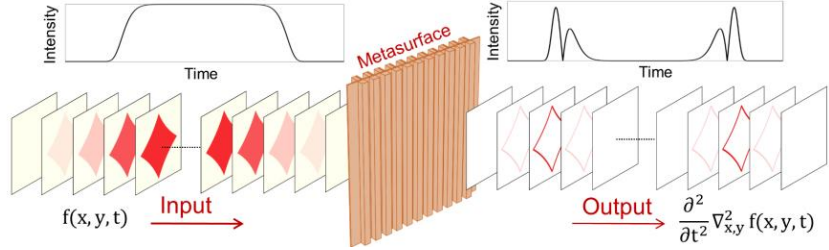

**Figure 1. Schematic of spatio-temporal edge-detection with metasurfaces.** Left: A spatio-temporal input signal $f(x, y, t)$, encoded in the envelope of a travelling wave, impinges on the metasurface (center). Right: The output signal corresponds to the space-time derivative of the input.

We begin by discussing how a general spatio-temporal differential operator can be implemented with nonlocal metasurfaces (MSs). For simplicity, we focus on 1D images, but the concept can be extended to 2D. A general time-varying image, carried by a structured electromagnetic wave with carrier frequency $\omega_0$, is described by $\tilde{f}(x,t) = e^{i\omega_0 t} f(x,t) = e^{i\omega_0 t} \int d\Omega \, dk_x e^{-i[k_x x - \Omega t]} F(k_x, \Omega)$, where $f(x,t)$ is a slowly-varying (in time) envelope function, $F(k_x, \Omega)$ is its Fourier transform, and $\Omega \equiv \omega - \omega_0$. When the image impinges on a linear MS, the transmitted image is determined by the filtering of spatio-temporal Fourier components, i.e., by the MS momentum-frequency transfer function (TF) $t(k_x, \Omega)$. In particular, a mixed spatio-temporal derivative $\frac{\partial^m}{\partial x^m} \frac{\partial^n}{\partial t^n} f(x,t)$ is obtained if the TF is $t^{(m,n)}(k_x, \Omega) = k_x^m \Omega^n$ (up to an overall phase factor). Because the second-order differentiation is better suited for isotropic edge detection[16,27], here we focus on the operator $\frac{\partial^2}{\partial x^2}\frac{\partial^2}{\partial t^2}$, but our approach can be generalized to other operators. Our target response corresponds to the TF $t^{(2,2)}(k_x,\Omega) = k_x^2\Omega^2$ (Fig. 2c), which is achieved when a MS supports a transmission zero at $\Omega = 0$ and normal-incidence ($k_x = 0$) and, additionally, the transmission increases only when *both* $\Omega$ and $k_x$ are simultaneously varied, while it remains zero when either $\Omega$ or $k_x$ is zero. Such non-trivial TF is markedly different from the one necessary to implement linear combinations of spatial and temporal derivatives[34–37]. Designing MSs with arbitrary spatio-temporal nonlocal TFs is generally nontrivial. Inverse-design methods can help this task[4,32], and spatio-temporal coupled-mode theories can provide powerful guidelines[37,44].

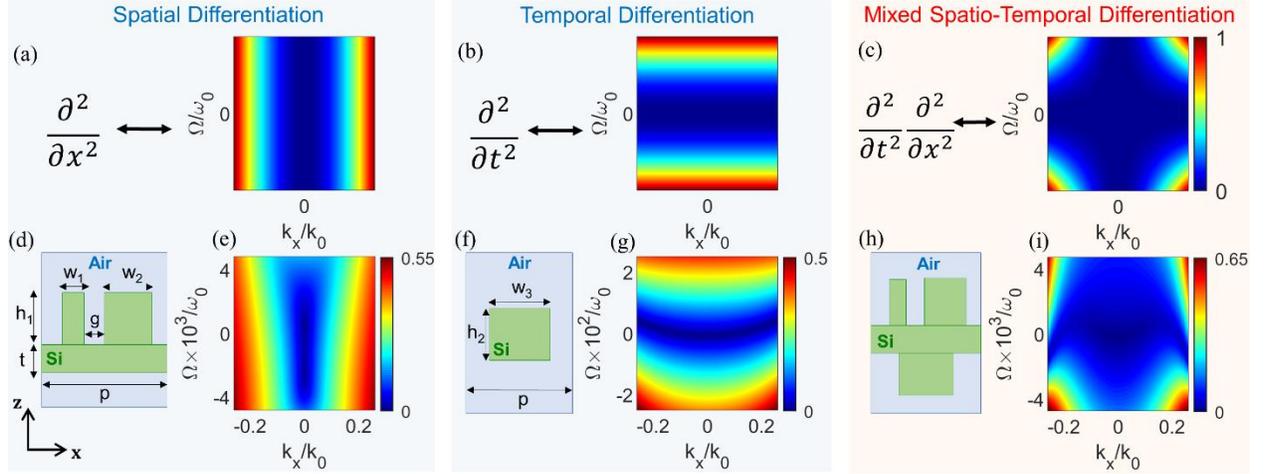

**Figure 2. Ideal transfer function (TF), geometry and numerical simulations of the proposed MS.** (a-c) Magnitude of the ideal TF to perform (a) second-order spatial derivative, (b) second-order time derivative, and (c) mixed spatio-temporal derivative. (d-i) Unit cells and numerically calculated TFs of realistic metasurfaces performing the corresponding operations shown on top of each box. The parameters are $p = 810\ nm$, $h_2 = 393\ nm$, $w_3 = 430\ nm$, $h_1 = 540\ nm$, $w_1 = 80\ nm$, $w_2 = 260\ nm$, $g = 55\ nm$ and $t = 83$ nm. The geometrical dimensions of the metasurface in panel h are provided in [45].

Here, we use a more intuitive approach, whereby the required spatio-temporal response is obtained by engineering the spatial and temporal nonlocalities of two cascaded MSs. Indeed, the TF $t^{(2,2)}(k_x,\Omega)$ can be realized as the product of the TF of a MS performing spatial-only differentiation $[(t^{(2,0)}(k_x,\Omega) = k_x^2$, Fig. 2a] and the TF of a MS performing temporal-only differentiation $[(t^{(0,2)}(k_x,\Omega) = \Omega^2$, Fig. 2b]. Following this intuition, we optimize two different MSs that separately perform these two operations[14,33]. For concreteness, we assume silicon-based 1D MSs embedded in air (Fig. 2), and we target an operational wavelength $\lambda \approx 1500$ nm ($\frac{\omega_0}{2\pi} \approx 200$ THz ). All electromagnetic simulations in Fig. 2 were performed with Comsol. The MSs are orthogonal to the z direction and uniform along y, and the impinging electric field is y-polarized. Fig. 2d shows the geometry of a 1D MS whose TF, $t_s^{MS}(k_x,\Omega)$, closely approximates the TF required for second-order spatial derivative, $t^{(2,0)}(k_x,\Omega)$ (see figure caption for the geometry). We note that the design in Fig. 2d allows breaking horizontal and/or vertical mirror symmetries. Such broken symmetries are not fundamentally required to implement the even operator $\frac{\partial^2}{\partial x^2}$, but they would be required for odd-order derivatives[46,47]. Fig. 2f, instead, shows the unit cell of another MS whose TF, $t_t^{MS}(k_x, \Omega)$, is close to the one implementing second-order temporal differentiation [$t^{(0,2)}(k_x, \Omega)$]. In first approximation, i.e., neglecting near-field perturbations and feedback due to back-and-forth reflections, cascading these two MSs leads to a device with a TF given by the product $t_s^{MS}(k_x, \Omega) \times t_t^{MS}(k_x, \Omega)$, realizing the desired spatio-temporal derivation. In practice, multiple reflections are non-negligible, especially in this scenario where each individual MS is highly reflective at normal incidence. Nonetheless, since the two MS independently target nonlocality in orthogonal spaces, and they are designed to be non-dispersive in the dual coordinate, cascading the two designs shown in Figs. 2(d,f) provides a useful first-order design to obtain the desired functionality (see also [45]). The residual feedback due to multiple reflections is minimized by adjusting the distance between the MSs to prevent resonances, and by finely tuning the geometrical parameters to account for spectral shifts induced by feedback. Remarkably, for this particular device we found that, even when the two MSs are cascaded back-to-back without any gap between them, after fine tuning their geometry, the resulting metasurface (Fig. 2h) supports a TF [denoted $t_{st}^{MS}(k_x, \Omega)$, Fig. 2i] that agrees well with the target one (Fig. 2c). However, we emphasize that such zero-gap configuration might not apply to other devices and functionalities. The phases of all TFs shown in Fig. 2 are reported in [45].

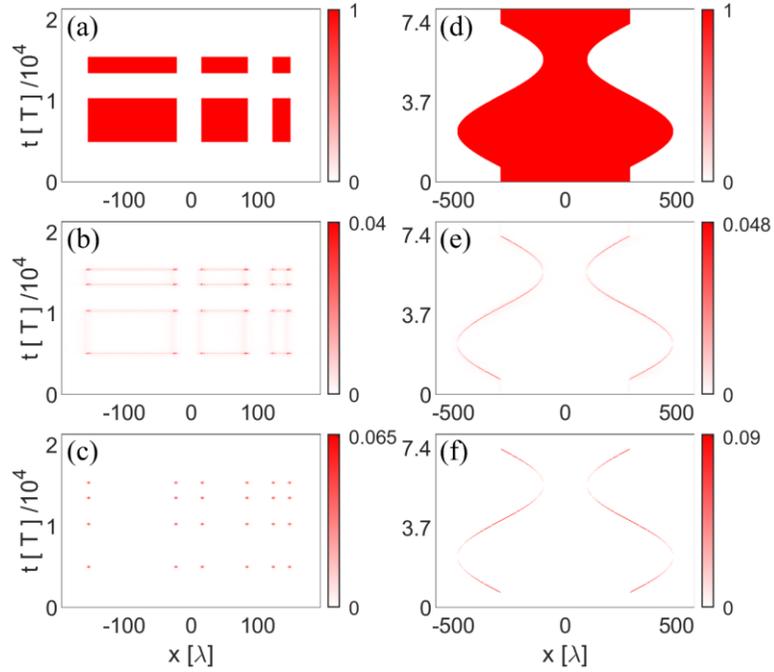

**Figure 3. Spatio-temporal intensity of input and output signals**. (a) Time-dependent 1D image consisting of segments with fixed size and whose intensity is switched on and off in time. (b-c) Corresponding output image, calculated using (b) the realistic transfer function (Fig. 2i) and (c) the ideal transfer function (Fig. 2c). (d-f) Same as in panels a-c, but with a different spatio-temporal input, consisting of a single segment whose length changes in time. Times are in units of $T = \frac{2\pi}{\omega_0} \approx 4.5\ fs$ and lengths are in units of $\lambda \approx 1.5\ \mu m$.

While we were able to design a single-layer MS (Fig. 2h) with a TF (Fig. 2i) close to ideal (Fig. 2c), small discrepancies arise. In particular, $t_{st}^{MS}(k_x, \Omega)$ is not symmetric along the frequency axis, e.g., the high-transmission lobes at high frequencies are less pronounced than the ones at lower frequencies. In order to assess these discrepancies more quantitatively, we expanded the calculated TF $t_{st}^{MS}(k_x, \Omega)$ (Fig. 2i) over the basis $t^{(m,n)}(k_x, \Omega)$. This analysis, discussed in [45], reveals that the weight of the term $t^{(2,2)}(k_x, \Omega)$ in the expansion of $t_{st}^{MS}(k_x, \Omega)$ is larger than 93% of the combined total weights, confirming that $t_{st}^{MS}(k_x, \Omega)$ mainly performs the $\frac{\partial^2}{\partial x^2} \frac{\partial^2}{\partial t^2}$ operation. Moreover, the desired response is obtained over a broad range of wavevectors and frequencies: The spatial numerical aperture (NA$_s$), which defines the maximum spatial frequency of the input image that the MS can process, is NA$_s$ = k$_{x,max}$/k$_0$ = sin($\theta_{max}$) ≈ 0.26 ($\theta_{max}$ ≈ 15°) , corresponding to a maximum spatial resolution of R$_s$ = $\frac{\lambda}{2NA_s}$ ≈ 3 µm. Similarly, we define a temporal numerical aperture, NA$_t$ = $\Omega_{max}/\omega_0$ ≈ 0.005, which quantifies the maximum spectral bandwidth of the input signal. The metasurface temporal resolution (i.e., the smallest time separation that can be resolved in the input signal), is R$_t$ = π/($\omega_0$ NA$_t$) ≈ 0.5 ps. The ratio of these two numerical apertures defines the *characteristic speed* v$_0$ ≡ c NA$_t$/NA$_s$ (where c is the speed of light), an important parameter of the device discussed more below.

We now show how this MS can effectively implement spatio-temporal edge detection. We first consider the case where an input image composed of spatial segments of different lengths, whose intensities are abruptly switched between zero and one (Fig. 3a) is processed by the metasurface in Figs. 2(h-i). The spatial and temporal dimensions of the input are adjusted to fit the maximum spatial (NA$_s$) and temporal (NA$_t$) frequencies of the designed MS, respectively. As expected, in the calculated [45] output image (Fig. 3b) only the *spatio-temporal edges* – i.e., the areas where the input image features strong gradients both in space and time, corresponding to the corners of the rectangles in Fig. 3a – are largely enhanced with respect to other areas of the signal. In other words, the MS highlights spatial edges only when they also evolve in time. For comparison, Fig. 3c shows the output image calculated using the ideal TF (Fig. 2c), which exactly implements the $\frac{\partial^2}{\partial x^2} \frac{\partial^2}{\partial t^2}$ operation. The close resemblance between Figs. 3b and 3c confirms that the MS mainly performs the targeted spatio-temporal differentiation. Minor detrimental effects are also visible in Fig. 3b: the spatial-only or temporal-only edges are partially transmitted (although with weaker intensity than the spatio-temporal edges), due to nonidealities of the TF in Fig. 2i [45].

The designed MS also features a high throughput efficiency: the peak intensity transmitted in the metasurface-filtered image (Fig. 3b) is over 60% of the intensity obtained in the ideal scenario (Fig. 3c), comparable to the best efficiencies experimentally achieved in spatial-only edge detection[16]. Next, we show that our MS can also process input signals that are not factorized in spatial and temporal parts. In Fig.

3d, we consider a single segment with fixed intensity and whose width changes in time. As expected, in the filtered output image (Fig. 3e) the edges of the segment are enhanced only at times when the segment is undergoing a width change – i.e., when the effective position of a spatial edge is changing in time. Once again, the output image filtered by the metasurface (Fig. 3e) perfectly matches the image obtained by applying the exact spatio-temporal differentiation (Fig. 3f).

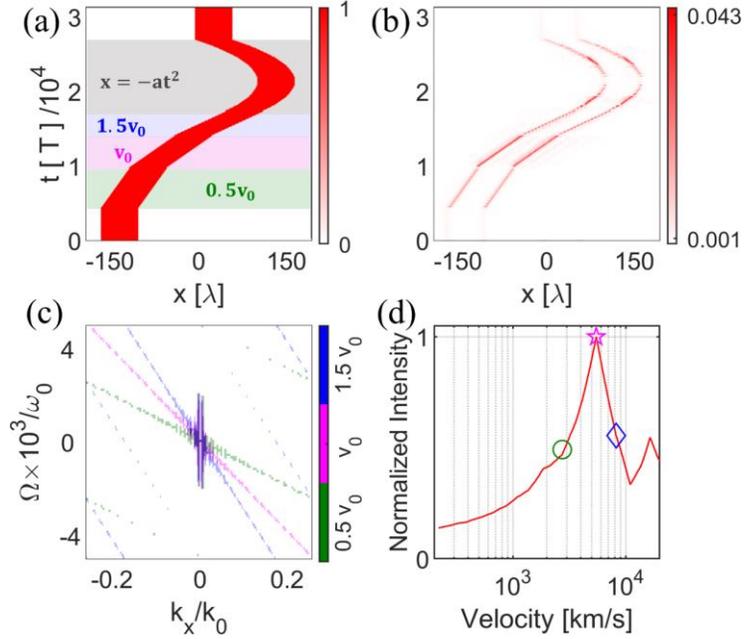

**Figure. 4. Spatio-temporal edge detection of a moving object**. (a) Time-dependent 1D image consisting of a segment with fixed width. The segment, initially at rest for a certain interval of time, starts moving with constant velocities 0.5 $v_0$ (green), $v_0$ (purple) and 1.5 $v_0$ (blue); then it negatively accelerates, and finally it stops. (b) Intensity of the output signal, highlighting the spatiotemporal edges. (c) Spatio-temporal Fourier Transform of the signals corresponding to objects moving uniformly with three different speeds (see text for details). (d) Intensity of the spatio-temporal edges versus the speed of the segment.

The proposed MS can be used for event-based edge detection – i.e., to enhance the edges of an object only when it is moving. In Fig. 4a, we consider a segment with fixed width which is initially at rest and then starts moving. In the first part of the motion, the object maintains a constant speed equal to 0.5 $v_0$, where $v_0$ is the characteristic speed defined above. The speed then increases in discrete steps, first to $v_0$ and then to 1.5 $v_0$, as indicated by the color-coded shaded areas in Fig. 4a. Then, it undergoes a uniformly accelerated motion (grey shaded area in Fig. 4a), until it stops again. In the filtered spatio-temporal signal (Fig. 4b), the edges of the object are enhanced only when the object is moving, as expected. Moreover, the edge intensities clearly depend on the object speed, with the largest intensity achieved when the object is moving at a speed close to $v_0$. This effect can be explained by considering how the spatio-temporal Fourier transform of a uniformly moving 1D object, $F_v(k_x, \Omega)$, depends on the object speed v. In Fig. 4c we consider an object (with the same size as in Fig. 4a) moving uniformly with velocities v = 0.5 $v_0$ (green), v = $v_0$ (purple) and v = 1.5 $v_0$ (blue). In order to highlight regions where each Fourier transform $F_v(k_x, \Omega)$ is largely different from zero, we associate to each speed v a binary function which is 0 when $|F_v(k_x, \Omega)|$ is smaller than 1.5% of its absolute maximum and 1 otherwise. By comparing Fig. 4c with Fig. 2i, it becomes clear that the spatio-temporal Fourier transform $F_v(k_x, \Omega)$ for an object moving with velocity $v = v_0$ (purple color) optimally aligns with the high transmission lobes of the transfer function, thus leading to higher intensities of the spatio-temporal edges. This behavior is also summarized in Fig. 4d, which shows the intensity of the spatio-temporal edges of the object versus the object speed. As expected, the intensity peaks at $v_0 = c\,NA_t/NA_s \approx 5800$ km/s (denoted by the purple star), and it decreases when the speed is either lower or higher than this optimal value. While the speed $v_0$ that maximizes the output intensity is fairly high, we note that the dependence of the output intensity on object speed (Fig. 4a) is weak, and reasonable output intensities (~20% of the peak) are still obtained for speeds that are over one order of magnitude smaller than $v_0$. Generally, for objects moving at velocity $v < v_0$ and detected via the operation $\partial_t^m \partial_x^n$, the intensity of the spatio-temporal edges approximately scales as $v^{2m}$. Moreover, $v_0$ can be tailored through the MS nonlocality. The optimal speed $v_0$ is determined by $NA_t$ and $NA_s$, and it can be controlled by varying them. For a fixed temporal numerical aperture $NA_t$, $v_0$ can be decreased by increasing the spatial numerical aperture $NA_s$, which would also increase the spatial resolution. For the device considered here ($NA_s = 0.26$), for example, the characteristic speed $v_0$ may be decreased by a factor of about 4 by working with larger $NA_s \to 1$. Alternatively, for a fixed spatial numerical aperture, $v_0$ can be reduced by reducing the temporal numerical aperture $NA_t$, linked to the spectral bandwidth of the MS performing the temporal differentiation (Fig. 2b, 2f-g). In particular, $v_0$ can be largely reduced by increasing the quality factor of the MS in Fig. 2(f-g). As an example, in [45] we show alternative designs with $v_0 \approx 3{,}000$ km/s and $v_0 \approx 1825$ km/s.

Here, we have extended the concept of nonlocality engineering to the space-time domain, demonstrating a passive nonlocal metasurface that can perform analog mixed spatio-temporal differentiation on an input image. This functionality can be used to perform event-based edge detection, whereby the edges of an input image are enhanced only when the image intensity varies in time. Such event-based analog processing necessitates mixed spatio-temporal differentiation, and it cannot be achieved with other forms of spatio-temporal devices[34–37]. This nontrivial functionality – which may replace bulky circuit-based neuromorphic cameras – is achieved with a subwavelength silicon metasurface, whose design is compatible with conventional fabrication methods. We have shown that this MS can be used to detect the edges of an image whose intensity evolves in time, or to enhance the edges of a moving object. In this latter case, the intensity of the filtered image depends on the object speed, which opens interesting applications in the fields of sensing and vibration monitoring. Moreover, we have elucidated how the metasurface design can be tailored to maximally enhance objects moving at desired speeds. The MS design can be readily extended to 2D image processing, paving the way towards optical analog event-based vision, leading to a vast spectrum of applications in fully-passive, low-energy, and ultrafast optical computing.

This work was supported by the Air Force Office of Scientific Research and the Simons Foundation.

# Supplementary Information for

# Space-Time Nonlocal Metasurfaces for Event-Based Image Processing

Sedigheh Esfahani[1,2,†], Michele Cotrufo[1,3,†], and Andrea Alù[1,2,*]

[1]*Photonics Initiative, Advanced Science Research Center, City University of New York, New York, NY 10031, USA*

[2]*Physics Program, Graduate Center of the City University of New York, New York, NY 10016, USA*

[3]*The Institute of Optics, University of Rochester, Rochester, New York, NY 14627, USA*

[†] *These authors contributed equally*

*\*aalu@gc.cuny.edu*

## S.1 Cascading the two metasurfaces

The metasurface design in Fig. 2h-i of the main text was obtained by cascading two metasurfaces that were separately optimized to feature the transfer functions $t^{(2,0)}(k_x,\Omega)$ (Fig. 2d-e) and $t^{(0,2)}(k_x,\Omega)$ (Fig.2f-g), respectively, followed by a fine optimization of the geometry to reduce detrimental effects due to Fabry-Perot-like resonances and near-field interactions. The gap between the two metasurfaces play a dominant role in this regard. To showcase this, in Fig. S1 we show the amplitude and phase of the transfer function of the cascaded device when the geometrical parameters of each metasurface are kept equal to the ones found in the optimization of each device (Fig. 2d and 2f of the main text), while the vertical gap between them is varied between 1 µm and 3 µm. For a finite gap, we found that the best performance is achieved when the gap is about 2500 nm (corresponding to $1.7\lambda_0$). We then investigated the possibility of reducing the gap to much smaller values. Remarkably, we found that when the gap is reduced to zero (Fig. 2h), the transfer function of the resulting metasurface (Fig. 2i) displays a good agreement with the target one. In the main text we displayed the amplitudes of the transfer functions in Figs. 2e and 2g and 2i. The panels in Fig. S2 show the corresponding phases.

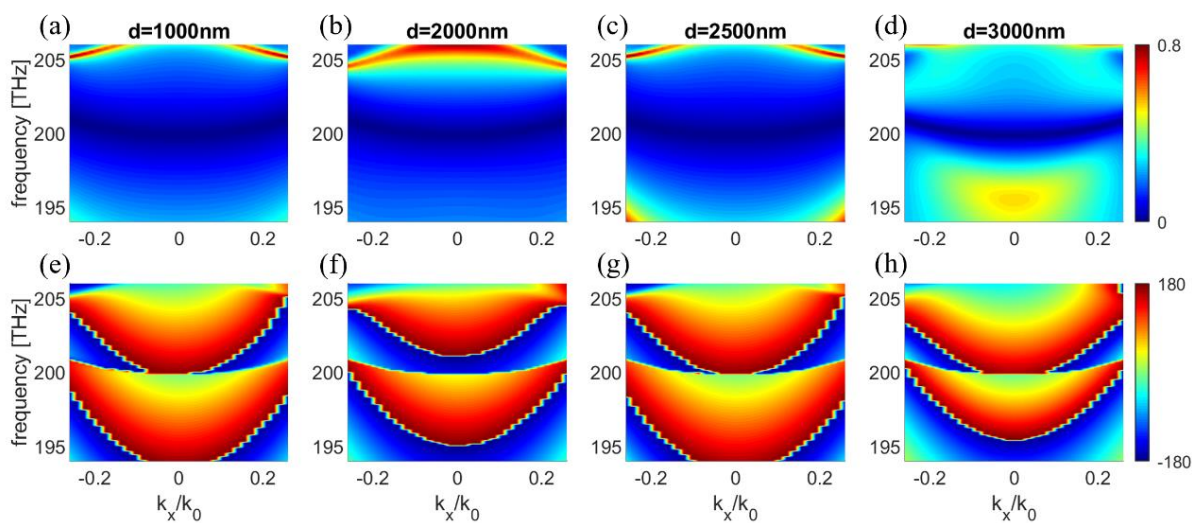

**Figure S1. Transfer functions of the cascaded device, from by the two MSs shown in Fig. 2d and 2e of the main text, as a function of vertical gap d between the MSs.** (a-d) Magnitudes of the TFs, (e-h) Phase of the TFs.

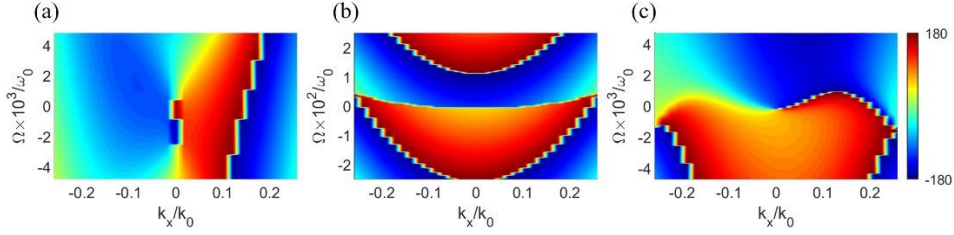

**Figure S2. Corresponding phases of Fig. 2e and 2g and 2i**. (a) Phase of the TF of the MS shown in Fig. 2e performing second-order spatial differentiation. (b) phase of the TF of the MS shown in Fig. 2g performing second-order temporal differentiation. (c) phase of the TF of the back-to-back cascaded MS shown in Fig. 2i performing spatio-temporal differentiation.

## S.2 Mathematical model

In order to assess more quantitatively the computational behavior of the optimized MS (Fig. 2h-i), we assume that the corresponding TF in Fig. 2i can be expanded as a sum of different terms, each corresponding to a combination of time and/or spatial differential operations, i.e., $t_{\text{Ansatz}}(k_x, \Omega) = \sum_{m,n} A_{m,n} \left(\frac{k_x}{k_0}\right)^m \left(\frac{\Omega}{\omega_0}\right)^n$, $m, n \in \mathbb{N}_0$, where $A_{m,n}$ are complex coefficients and, for numerical convenience, we normalized the frequencies and wavevectors by $\omega_0 = 2\pi \times 211.7$ THz and $k_0 = c/\omega_0$, respectively. We then use this ansatz to fit the numerically calculated complex TF in Fig. 2i, and then use the weights $\rho_{m,n} \equiv |A_{m,n}|/\sum_{r,s}|A_{r,s}|$ to quantify the contribution of each differential operator. To keep the analysis simple, we considered only values of $m, n = 0,1,2$. By performing this analysis, we find $\rho_{2,2} = 0.935$, confirming that the operation performed by this MS is very close to the targeted mixed differentiation $\frac{\partial^2}{\partial x^2}\frac{\partial^2}{\partial t^2} f(x,t)$. This metric further increases when considering all possible mixed differential operations, i.e., all terms in $t_{\text{Ansatz}}(k_x, \Omega)$ for which $m \cdot n \neq 0$. In particular, we find that $\sum_{(m \cdot n \neq 0)} \rho_{mn} > 0.98$. The remaining terms, contributing to the output with a weight $\sum_{(m \cdot n = 0)} \rho_{mn} < 0.02$, will introduce a small background in the filtered image, which is visible for example in Fig. 3b.

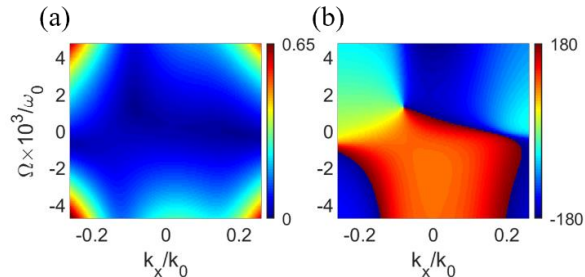

**Figure S3.** Mathematical fit of the TF in Fig. 2i considering the first nine terms in $t_{Ansatz}(k_x, \Omega) = \sum_{m,n} A_{m,n} \left(\frac{k_x}{k_0}\right)^m \left(\frac{\Omega}{\omega_0}\right)^n$, $m, n \in \mathbb{N}_0$. (a) Magnitude of $t_{Ansatz}(k_x, \Omega)$. (b) phase of $t_{Ansatz}(k_x, \Omega)$.

For completeness, in Fig. S3 we show the best-fit obtained by modeling the TF in in Fig. 2i with the function $t_{\text{Ansatz}}(k_x, \Omega) = \sum_{m,n} A_{m,n} \left(\frac{k_x}{k_0}\right)^m \left(\frac{\Omega}{\omega_0}\right)^n, m,n = 0,1,2$. The good agreement between the plots in Fig. S3 and amplitude (Fig. 2i) and the phase (Fig. S2c) of the TF of the proposed MS confirms that the Ansatz can correctly capture the main features of the TF.

## S.3 Numerical calculations of the filtered images

In this section we briefly explain how we performed the numerical calculations of the filtered images in Figs. 3 and 4 of the main text. As explained in the main text, the input image can be decomposed into its spatio-temporal Fourier components via $\tilde{f}(x,t) = e^{i\omega_0 t} f(x,t) = e^{i\omega_0 t} \int d\Omega \, dk_x \, e^{-i[k_x x - \Omega t]} F(k_x, \Omega)$. The output image can be then obtained by multyplying the Fourier transform $F(k_x, \Omega)$ by the desired transfer function, and then performing an inverse Fourier transform. In the numerical calculations, the spatial and temporal sizes of the pixels of the input signal are adjusted to match the spatial ($NA_s$) and temporal ($NA_t$) numerical apertures The spatio-temporal discrete Fourier and anti-Fourier transforms are performed with Matlab. In order to emulate a metasurface that performs the exact operation $\frac{\partial^2}{\partial x^2}\frac{\partial^2}{\partial t^2}$ (Figs. 3c and 3f), we defined the ideal TF $t^{\text{ideal}}(k_x, \Omega) = \left(\frac{k_x}{k_0 NA_s}\right)^2 \left(\frac{\Omega}{\omega_0 NA_t}\right)^2$, whose amplitude reaches a value of 1 at the corners of the domain, as shown in Fig. 2c.

## S.4 Manipulating the characteristic speed of the MS

Here, we show how by increasing the quality factor of the MS in Fig. 2d-e we can control the temporal numerical aperture $NA_t$, and consequently the characteristic speed $v_0$ of the cascaded space-time MS.

The quality factor of the resonant grating performing temporal differentiation (Fig. 2d-e) can be easily controlled by tuning the height and width of the dielectric beams[1]. In Fig. S3a we show the amplitude of the TF for the same configuration of Fig. 2f, with design parameters $h_2 = 423 \text{ nm}, w_3 = 415 \text{ nm}, p = 810 \text{ nm}$, Comparing Fig. S3a and Fig. 2g, we can see that the spectral bandwidth of the TF is nearly reduced to half. We then use these new design parameters $h_2$ and $w_3$ in the cascaded metasurface (Fig. 2h) and fine tune the geometry to target the ideal TF of Fig. 2c. In Fig. S3b and S3c we show the amplitude and phase of the resulting TF. The corresponding optimized design parameters are $p = 810 \text{ nm}, h_2 = 423 \text{ nm}, w_3 = 415 \text{ nm}, h_1 = 490 \text{ nm}, w_1 = 80 \text{ nm}, w_2 = 260 \text{ nm}, g = 44.5 \text{ nm}$ and $t = 85 \text{ nm}$. The corresponding numerical aperture is $NA_s = k_{x,max}/k_0 = \sin(\theta_{max}) \approx 0.24$ ($\theta_{max} \approx 14°$) and the temporal numerical aperture, $NA_t = \Omega_{max}/\omega_0 \approx 0.0023$. Therefore, the characteristic speed of this MS is reduced to $v_0 \approx 3{,}000$ km/s. The characteristic speed can be further reduced by further increasing the Q factor of the time-differentiating MS. In Fig. S3d-f we show the results of another design operating at 215.49 THz, with parameters $p = 810$ nm, $h_2 = 445$ nm, $w_3 = 404$ nm, $h_1 = 460$ nm, $w_1 = 80$ nm, $w_2 = 260$ nm, $g = 44.5$ nm and $t = 85$ nm. For this design, the spatial and temporal numerical apertures are $NA_s = 0.19$ and $NA_t = 0.0012$ respectively, and $v_0 \approx 1{,}850$ km/s.

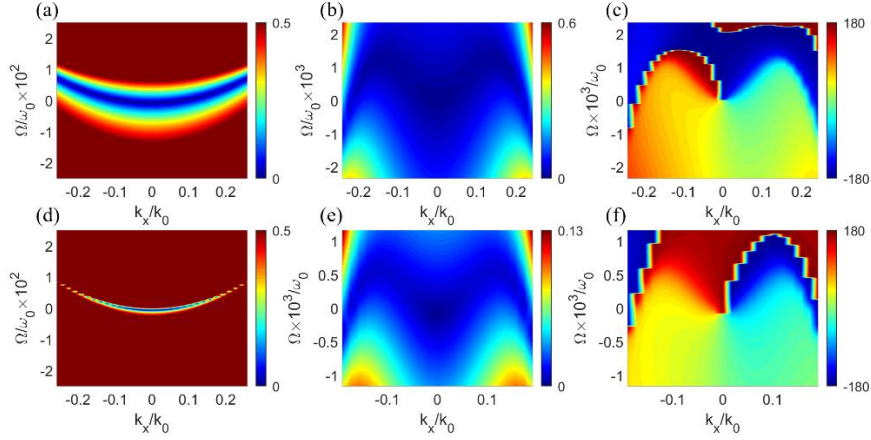

**Figure S4. Alternative designs of cascaded MSs with reduced characteristic speeds.** See text for details

## S.5 Unwanted background level due to nonidealities in the transfer function

In Fig. 3 of the main text, we have considered the case of a 1d image (a collection of segments) whose intensity changes over time. We have calculated the corresponding output image assuming that the input image is processed either by an metasurface with an ideal transfer function which exactly implements the $\frac{\partial^2}{\partial x^2}\frac{\partial^2}{\partial t^2}$ operation (Fig. 3b), or by a metasurface with the realistic transfer function shown in Fig. 2i (Fig. 3c). The results are reproduced in Figs. S5a and S5b, respectively. In the case where the ideal transfer function is considered (Fig. S5b), only the spatio-temporal edges (that is, the corners of the spatio-temporal rectangles in the input image) are enhanced, while any other signal in the input image is suppressed. In the case where the realistic transfer function is used (Fig. S5a), the results are similar, but with some minor detrimental effects: the spatial-only or temporal-only edges are partially transmitted (although with weaker intensity than the spatio-temporal edges), due to nonidealities of the transfer function in Fig. 2i of the main paper. This residual "unwanted" intensity is, however, much smaller than the intensity at the spatio-temporal edges. This is clearly shown in Fig. S5. Figures S5c and S5d show 1D slices of the output signals calculated with the ideal transfer function (dashed lines) and with the realistic metasurface transfer function (solid lines). The ideal case is characterized by narrow peaks in correspondence of the spatio-temporal edges. In the signal processed by the metasurface, we still obtain narrow peaks, and at the same positions as in the ideal case. However, we can also see some additional background signal between the peaks. Nonetheless, the signal-to-background ratio is always larger than 5, which allows to clearly discriminate the spatio-temporal edges from the spatial-only and temporal-only edges.

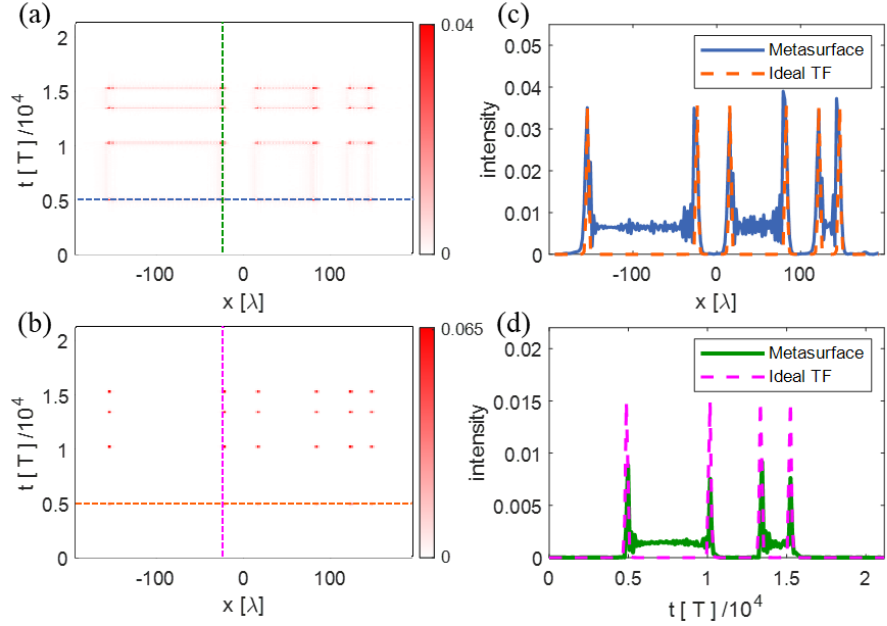

**Figure S5.** (a-b) Same plot as in Figs. 3(b-c) of the main text. (c-d) 1D slices extracted from panels a and b, corresponding to the color-coded dashed lines in panels a and b.

## S.6 Additional numerical data

In Fig. 2i of the main text we have shown the amplitude of the calculated transfer function of the final design (Fig. 2h). In Fig. S6 we show the same data but now on a logarithmic scale. It can be seen that at the center frequency and for normal incidence, the transmission amplitude remains below 0.001. The transmission amplitude also remains fairly low along the $k_x = 0$ and $\omega = \omega_0$ axis (which corresponds to the $\Omega = 0$ axis), which is the important requirement to implement the mixed derivative $\partial_x^2 \partial_t^2$. The fact that the transfer function is not exactly zero along the $k_x = 0$ and $\omega = \omega_0$ axis creates the unwanted background discussed in section S.5.

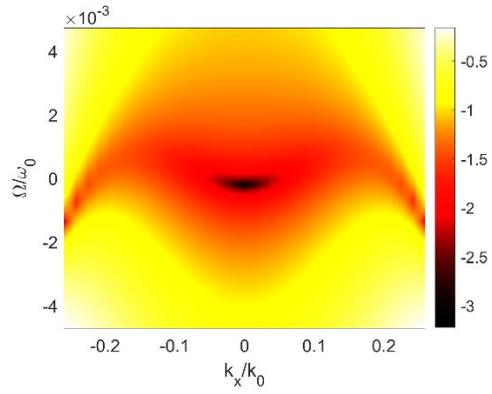

**Figure S6.** Same data as in Fig. 2i of the main text, but plotted in a logarithmic scale.